\documentclass[runningheads,a4paper]{llncs}

\usepackage{amssymb}
\usepackage{graphicx}
\usepackage{amsmath}
\usepackage{epsfig}
\usepackage{epstopdf}
\DeclareMathOperator{\trace}{tr}

\begin{document}

\mainmatter  % start of an individual contribution

\title{A General Framework for Online Audio Source Separation}

\titlerunning{A General Framework for Online Audio Source Separation}

\author{Laurent S. R. Simon \and Emmanuel Vincent}

\institute{INRIA, Centre de Rennes - Bretagne Atlantique\\
Campus de Beaulieu, 35042 Rennes Cedex, France\\
\{laurent.s.simon@inria.fr, emmanuel.vincent@inria.fr\}}

\toctitle{A General Framework for Online Audio Source Separation}
%\tocauthor{Laurent S. R. Simon \and Emmanuel Vincent}
\maketitle

\begin{abstract}
We consider the problem of online audio source separation. Existing algorithms adopt either a sliding block approach or a stochastic gradient approach, which is faster but less accurate. Also, they rely either on spatial cues or on spectral cues and cannot separate certain mixtures. In this paper, we design a general online audio source separation framework that combines both approaches and both types of cues. The model parameters are estimated in the Maximum Likelihood (ML) sense using a Generalised Expectation Maximisation (GEM) algorithm with multiplicative updates. The separation performance is evaluated as a function of the block size and the step size and compared to that of an offline algorithm.
\end{abstract}

\section{Introduction}
\label{sec:intro}

Audio source separation is the process of recovering a set of audio signals from a given mixture signal. This can be addressed via established approaches such as Independent Component Analysis (ICA), binary masking and Sparse Component Analysis (SCA) \cite{makino_blind_2007} or more recent approaches such as local Gaussian modeling and Nonnegative Matrix Factorisation (NMF) \cite{vincent_probabilistic_2010}. Most current algorithms are offline algorithms which require the whole signal in order to estimate the sources. In this paper, we focus on online audio source separation, whereby only the past samples of the mixture are available. This constraint arises in particular in real-time scenarios. 

A few online implementations have been designed for ICA \cite{mukai_real-time_2003} \cite{mori_blind_2006}, time-frequency masking \cite{loesch_online_2009}, local Gaussian modeling \cite{togami_online_2011}, spectral continuity-based separation \cite{ono_real-time_2008} and NMF \cite{wang_online_2011}. However, these algorithms rely either on spatial cues \cite{mukai_real-time_2003} -- \cite{togami_online_2011} or on spectral cues \cite{ono_real-time_2008,wang_online_2011} alone. Such algorithms are not capable of separating mixtures where several sources have the same spatial position and several sources have similar spectral characteristics. For example, in pop music, the voice, the snare drum, the bass drum and the bass are often mixed to the centre and several voices or several guitars are present.

In order to address this issue, we consider the general flexible source separation framework in \cite{ozerov_general_????}. This framework generalises a wide range of algorithms such as certain forms of ICA, local Gaussian modeling and NMF, and enables the specification of additional constraints on the sources such as harmonicity. By jointly exploiting spatial and spectral cues, it makes it possible to robustly separate difficult mixtures such as above. 

The two main approaches for online source separation are the sliding block (also known as blockwise) approach, as used in \cite{mukai_real-time_2003} \cite{mori_blind_2006} \cite{loesch_online_2009} \cite{ono_real-time_2008}, and the stochastic gradient (also known as stepwise) approach, as used in \cite{togami_online_2011} \cite{wang_online_2011}. The sliding block method consists in applying the offline audio source separation algorithm to a block of $M$ time frames. Once this block of signal has been processed, a frame is extracted for each of the $J$ sources before sliding the processing block by one frame. This approach is computationally intensive but accurate. The stepwise method offers to update the model parameters in every frame using only the latest available frame and the model parameters estimated in the previous frame. As it uses only the latest available frame at a given time, this approach is faster than the sliding block approach but can be inaccurate.

In this paper, we propose a general iterative online algorithm for the source separation framework in \cite{ozerov_general_????} that combines the sliding block approach and the stepwise approach using two hyper-parameters: the block size $M$ and the step size $\alpha$. As a by-product, we provide a way of circumventing the annealing procedure in \cite{ozerov_general_????}, which would require a large number of iterations per block. Moreover, we determine the best trade-off between these two approaches experimentally on a set of real-world music mixtures.

The structure of the rest of the paper is as follows: the flexible framework in \cite{ozerov_general_????} is introduced in Section 2. Section 3 presents the online algorithm. Experimental results are shown in Section 4. The conclusion can be found in Section 5. 

\section{General audio source separation framework}

We operate in the time-frequency (TF) domain by means of the Short-Time Fourier Transform (STFT). In each frequency bin $f$ and each time frame $n$, the multichannel mixture signal ${\bf x}(f,n)$ can be expressed as 
\begin{equation}
{\bf x}(f,n) = \sum_{j = 1}^J {\bf c}_j(f,n)
\end{equation}

\noindent where $J$ is the number of sources and ${\bf c}_j(f,n)$ is the STFT of the spatial image of the $j$-th source. 

\subsection{Model}

We assume that ${\bf c}_j(f,n)$ is a complex-valued Gaussian random vector with zero mean and covariance matrix ${\bf R}_{{\bf c}_j}(f,n)$
\begin{equation}
{\bf c}_j \sim \mathcal{N}_c({\bf 0}, {\bf R}_{{\bf c}_j})
\end{equation}

\noindent and that ${\bf R}_{{\bf c}_j}(f,n)$ factors as
\begin{equation}
{\bf R}_{{\bf c}_j}(f,n) = {\bf R}_{j}(f) v_j(f,n)
\end{equation}

\noindent where ${\bf R}_{j}(f)$ is the spatial covariance matrix of the $j$-th source and $v_j(f,n)$ is its spectral variance. 

In \cite{ozerov_general_????}, ${\bf R}_{j}(f)$ is expressed as ${\bf R}_{j}(f) = {\bf A}_j(f) {\bf A}_j^H(f)$, and ${\bf A}_j(f)$ is estimated instead. This results in an annealing procedure, which would translate into a large number of iterations within each block in our context. In order to circumvent the annealing, we assume that ${\bf R}_{j}(f)$ is full-rank and directly estimate ${\bf R}_{j}(f)$ instead, similarly to \cite{duong_under-determined_2010}.

The spectral variance $v_j(f,n)$ is modeled via a form of hierarchical NMF \cite{ozerov_general_????}. The matrix of spectral  variances ${\bf V}_j \triangleq [v_j(f,n)]_{f,n}$ is first decomposed into the product of an excitation spectral power ${\bf V}_j^{\textrm{x}}$ and a filter spectral power ${\bf V}_j^{\textrm{f}}$
\begin{equation}
{\bf V}_j = {\bf V}_j^{\textrm{x}} \odot {\bf V}_j^{\textrm{f}}
\end{equation}

\noindent where $\odot$ denotes entrywise multiplication. ${\bf V}_j^{\textrm{x}}$ is further decomposed into the product of a matrix of narrowband spectral patterns  ${\bf W}_j^{\textrm{x}}$, a matrix of spectral envelope weights ${\bf U}_j^{\textrm{x}}$, a matrix of temporal envelope weights ${\bf G}_j^{\textrm{x}}$ and a matrix of time-localised temporal patterns ${\bf H}_j^{\texttt{x}}$, so that
\begin{equation}
{\bf V}_j^{\texttt{x}} = {\bf W}_j^{\textrm{x}} {\bf U}_j^{\textrm{x}} {\bf G}_j^{\textrm{x}} {\bf H}_j^{\textrm{x}}.
\end{equation}

\noindent ${\bf V}_j^{\texttt{f}}$ is decomposed in a similar way.

This factorisation enables the specification of various spectral or temporal constraints over the sources. For example, harmonicity can be enforced by fixing ${\bf W}_j^{\textrm{x}}$ to a set of narrowband harmonic patterns.

\subsection{Offline EM-MU algorithm}
\label{sec:offline}

In an offline context, the model parameters are estimated in the Maximum Likelihood (ML) sense by a Generalised Expectation-Maximi\-sation (GEM) algorithm combined with Multiplicative Updates (MU) applied to the complete data $\{{\bf c}_j(f,n) \}$.

The log-likelihood is defined using the empirical mixture covariance matrix ${\bf \widehat R}_{{\bf x}}(f,n)$ \cite{duong_under-determined_2010} as
\begin{equation}
\log \mathcal{L} = \sum_{f,n} -\trace \big({\bf R}_{\bf x}^{-1}(f,n) {\bf \widehat R}_{\bf x}(f,n)\big) - \log \det (\pi {\bf R}_{\bf x}(f,n))
\end{equation} 

\noindent where 
\begin{equation}
{\bf R}_{\bf x}(f,n) = \sum_{j = 1}^J {\bf R}_{{\bf c}_j}(f,n)
\end{equation} 
\noindent is the covariance of the mixture ${\bf x}(f,n)$.

In the E-step, the expectation of the natural statistics is computed via \cite{duong_under-determined_2010}
\begin{eqnarray}\label{Omega}
{\bf \Omega}_j(f,n) & = &{\bf R}_{{\bf c}_j} (f,n) {\bf R_x^{-1}} (f,n) \\\label{hat_Rc}
{\bf \widehat R}_{{\bf c}_j}(f,n) & = & {\bf \Omega}_j(f,n) {\bf \widehat R}_{\bf x}(f,n){\bf \Omega}_j^H(f,n) + ({\bf I} - {\bf W}_j(f,n)){\bf R}_{{\bf c}_j}(f,n)
\end{eqnarray}

\noindent where ${\bf \Omega}_j$ is the Wiener filter, $\bf I$ is the $I \times I$ identity matrix and $I$ is the number of channels of the mixture.

In the M-step, the model parameters are updated as \cite{ozerov_general_????,duong_under-determined_2010}
\begin{eqnarray}
&{\bf R}_j (f) & = \frac{1}{N}\sum_{n=1}^N \frac{1}{v_j(f,n)}{\bf \widehat R}_{{\bf c}_j} (f,n)\\
&{\bf W}_j^{\textrm{x}} & = {\bf W}_j^{\textrm{x}} \odot \frac{[{\bf \widehat \Xi}_{j} \odot {{\bf V}_j^{\textrm{x}}}.^{-2} \odot {{\bf V}_j^{\textrm{f}}}.^{-1} ]({\bf U}_j^{\textrm{x}} {\bf G}_j^{\textrm{x}} {\bf H}_j^{\textrm{x}})^{T}}{ {\bf V}_j^{\textrm{x}}.^{-1}({\bf U}_j^{\textrm{x}} {\bf G}_j^{\textrm{x}} {\bf H}_j^{\textrm{x}})^{T}}\\
&{\bf U}_j^{\textrm{x}} & = {\bf U}_j^{\textrm{x}} \odot \frac{{{\bf W}_j^{\textrm{x}}}^T [{\bf \widehat \Xi}_{j} \odot {{\bf V}_j^{\textrm{x}}}.^{-2} \odot {{\bf V}_j^{\textrm{f}}}.^{-1} ]({\bf G}_j^{\textrm{x}} {\bf H}_j^{\textrm{x}})^{T}}{{{\bf W}_j^{\textrm{x}}}^T {\bf V}_j^{\textrm{x}}.^{-1}({\bf G}_j^{\textrm{x}} {\bf H}_j^{\textrm{x}})^{T}}\\\label{G}
&{\bf G}_j^{\textrm{x}} & = {\bf G}_j^{\textrm{x}} \odot \frac{({\bf W}_j^{\textrm{x}}{\bf U}_j^{\textrm{x}})^T [{\bf \widehat \Xi}_{j} \odot {{\bf V}_j^{\textrm{x}}}.^{-2} \odot {{\bf V}_j^{\textrm{f}}}.^{-1} ]{{\bf H}_j^{\textrm{x}}}^{T}}{({\bf W}_j^{\textrm{x}}{\bf U}_j^{\textrm{x}})^T {\bf V}_j^{\textrm{x}}.^{-1}{{\bf H}_j^{\textrm{x}}}^{T}}\\\label{H}
&{\bf H}_j^{\textrm{x}} & = {\bf H}_j^{\textrm{x}} \odot \frac{({\bf W}_j^{\textrm{x}} {\bf U}_j^{\textrm{x}} {\bf G}_j^{\textrm{x}})^T [{\bf \widehat \Xi}_{j} \odot {{\bf V}_j^{\textrm{x}}}.^{-2} \odot {{\bf V}_j^{\textrm{f}}}.^{-1} ]}{({\bf W}_j^{\textrm{x}} {\bf U}_j^{\textrm{x}} {\bf G}_j^{\textrm{x}})^T {\bf V}_j^{\textrm{x}}.^{-1}}
\end{eqnarray}
   
\noindent where $.^{p}$ denotes entrywise raising to the power $p$,  $N$ is the number of time frames in the STFT of the signal and ${\bf \widehat \Xi}_{j} = [\widehat \xi_{j}(f,n)]_{f,n}$, with
\begin{equation}
{\widehat \xi}_j(f,n) = \frac{1}{I}\trace({\bf R}_j^{-1} (f) {\bf \widehat R}_{{\bf c}_{j}} (f,n)).
\end{equation}

\noindent ${\bf W}_j^{\textrm{f}}$, ${\bf U}_j^{\textrm{f}}$, ${\bf G}_j^{\textrm{f}}$ and ${\bf H}_j^{\textrm{f}}$ are updated in a similar way.

After each EM iteration, the model parameters are normalised: the mean of ${\bf R}_j$, ${\bf W}_j^{\textrm{x}}$, ${\bf U}_j^{\textrm{x}}$, ${\bf G}_j^{\textrm{x}}$, ${\bf H}_j^{\textrm{x}}$, ${\bf W}_j^{\textrm{f}}$, ${\bf U}_j^{\textrm{f}}$ and ${\bf H}_j^{\textrm{f}}$ are normalised to 1 while ${\bf G}_j^{\textrm{f}}$ is multiplied by the product of the normalisation factors of the other variables.

The separated sources are then obtained via
\begin{equation}\label{eq:xi}
{\bf \widehat c}_j(f,n) = {\bf \Omega}_j(f,n){\bf x}(f,n).
\end{equation}

\section{Online EM-MU algorithm}

We now consider an online context where in each time frame $t$, the data is limited to a block of $M$ STFT frames indexed by $n$ with $t-M+1 \leq n \leq t$, where $M = 1$  for the stepwise approach and $M = N$ for the full offline approach. We define a step size coefficient $\alpha \in \,\, ]0; 1]$ to stabilise the parameter updates by averaging over time. For each block, the spatial covariance matrices ${\bf R}_j^{(t)} (f)$ are initialised to a diffuse spatial covariance spanning a part of the audio space. The temporal weights ${{\bf G}_j^{\textrm{x}}}^{(t)}$ are randomly initialised and the normalised to the mean spectral power of the signal. Finally, the temporal patterns ${{\bf H}_j^{\textrm{x}}}^{(t)}$ are initialised to diagonal matrices. The expectation of the natural statistics is computed using (\ref{Omega}) and (\ref{hat_Rc}) for $t-M+1 \leq n \leq t$, whilst the spatial covariance matrix is updated as follows:
\begin{eqnarray}	\label{eq:onlineRj}	
{\bf R}_j^{(t)} (f) & = &(1 - \alpha) {\bf R}_j^{(t-1)} (f) +  \alpha \left(\frac{1}{M}\sum_{n=t-M+1}^t \frac{1}{v_{j}(f,n)}{\bf \widehat R}_{{\bf c}_{j}} (f,n)\right)
\end{eqnarray}

\noindent where the superscript $^{(t)}$ denotes is the value of matrix for the block $t$.

${{\bf G}_j^{\textrm{x}}}^{(t)}$ and ${{\bf H}_j^{\textrm{x}}}^{(t)}$ are updated using (\ref{G}) and (\ref{H}) for $t-M+1 \leq n \leq t$, as they are expected to significantly vary between blocks, whereas the updates of ${\bf W}_j^{\textrm{x}}$ and ${\bf U}_j^{\textrm{x}}$ become
\begin{eqnarray}
{{\bf W}_j^{\textrm{x}}}^{(t)} & = &{{\bf W}_j^{\textrm{x}}}^{(t)} \odot \frac{{{\bf M}_j^{\textrm{x}}}^{(t)}}{{{\bf C}_j^{\textrm{x}}}^{(t)}}\\\label{eq:onlineU}
{{\bf U}_j^{\textrm{x}}}^{(t)} & = &{{\bf U}_j^{\textrm{x}}}^{(t)} \odot \frac{{{\bf N}_j^{\textrm{x}}}^{(t)}}{{{\bf D}_j^{\textrm{x}}}^{(t)}}
 \end{eqnarray}
	
\noindent where
\begin{eqnarray}
{{\bf M}_j^{\textrm{x}}}^{(t)} & = &(1-\alpha){{\bf M}_j^{\textrm{x}}}^{(t-1)} + \alpha[{\bf \widehat \Xi}_{j} \odot {{\bf V}_j^{\textrm{x}}}.^{-2} \odot {{\bf V}_j^{\textrm{f}}}.^{-1} ]({{\bf U}_j^{\textrm{x}}}^{(t)} {{\bf G}_j^{\textrm{x}}}^{(t)} {{\bf H}_j^{\textrm{x}}}^{(t)})^{T}\\
{{\bf C}_j^{\textrm{x}}}^{(t)} & = &(1-\alpha){{\bf C}_j^{\textrm{x}}}^{(t-1)} + \alpha{\bf V}_j^{\textrm{x}}.^{-1}({{\bf U}_j^{\textrm{x}}}^{(t)} {{\bf G}_j^{\textrm{x}}}^{(t)} {{\bf H}_j^{\textrm{x}}}^{(t)})^{T}\\
{{\bf N}_j^{\textrm{x}}}^{(t)} & = & (1-\alpha){{\bf N}_j^{\textrm{x}}}^{(t-1)} + \alpha{{\bf W}_j^{\textrm{x}}}^{(t)T} [{\bf \widehat \Xi}_{j} \odot {{\bf V}_j^{\textrm{x}}}.^{-2} \odot {{\bf V}_j^{\textrm{f}}}.^{-1} ]({{\bf G}_j^{\textrm{x}}}^{(t)} {{\bf H}_j^{\textrm{x}}}^{(t)})^{T}\\
{{\bf D}_j^{\textrm{x}}}^{(t)} & = & (1-\alpha){{\bf D}_j^{\textrm{x}}}^{(t-1)} + \alpha{{\bf W}_j^{\textrm{x}}}^{(t)T} {\bf V}_j^{\textrm{x}}.^{-1}({{\bf G}_j^{\textrm{x}}}^{(t)} {{\bf H}_j^{\textrm{x}}}^{(t)})^{T}
\end{eqnarray}

\noindent where ${\bf \widehat \Xi}_{j}^{(t)}$ is computed as in (\ref{eq:xi}). ${{\bf M}_j^{\textrm{f}}}^{(t)}$, ${{\bf C}_j^{\textrm{f}}}^{(t)}$, ${{\bf N}_j^{\textrm{f}}}^{(t)}$ and ${{\bf D}_j^{\textrm{f}}}^{(t)}$ are updated in a similar way. At each block, several iterations can be performed in order to improve the estimation of the model parameters.

Although equations (\ref{eq:onlineRj}) to (\ref{eq:onlineU}) look similar to the online update of the local Gaussian model in \cite{togami_online_2011} and \cite{wang_online_2011}, there are two crucial differences:
\begin{itemize}
\item The framework introduced in the current paper is more general in the sense that it uses hierarchical NMF, enabling the user to apply more specific constraints than when using shallow NMF.
\item It is not limited to the sole use of the latest audio frame.
\end{itemize}

\section{Experimental results}

We compared the performance of the online audio source separation framework to the offline framework introduced in section \ref{sec:offline}, as a function of the number of EM iterations, $\alpha$ and $M$. The project aiming at remixing of recordings for sound engineers, DJs and consumers, we processed five 10~s long stereo commercial pop recordings composed of bass, drums, guitars, strings and voice. All the recordings were recorded at 44100 Hz. The STFT was computed using half-overlapping 2048 sample sine windows. In the offline algorithm as well as in the online algorithm, each of the modeled sources were constrained in a way similar to section  V.\emph{C} in \cite{ozerov_general_????}. In the case of an harmonic source, ${{\bf W}_j^{\textrm{x}}}^{(t)}$ was fixed to a set of narrowband harmonic spectral patterns and the spectral envelope weights in ${{\bf U}_j^{\textrm{x}}}^{(t)}$ were updated, whereas for bass and percussive sources, ${{\bf W}_j^{\textrm{x}}}^{(t)}$ was a fixed diagonal matrix and ${{\bf U}_j^{\textrm{x}}}^{(t)}$ was a fixed matrix of basis spectra learned over a corpus of bass and drum sounds.

Audio samples of the separated sounds of this experiment can be found on \\http://www.irisa.fr/metiss/lssimon/LVA2012/index.html .

Separation performance was evaluated using the Signal-to-Distortion Ratio (SDR), the Signal-to-Interference Ratio (SIR), the source Image to Spatial distortion Ratio (ISR) and the Source-to-Artifacts Ratio (SAR) defined in \cite{vincent_first_2007}. For each set of conditions over the number of iterations, $M$ and $\alpha$, each of these criteria was averaged over all the mixtures and all the separated sound sources. Over all the results of this experiment, the SDR varied between -1.1 and 0.9 dB, the SIR between -4 and 1 dB, the ISR between 2.3 and 3.9 dB and the SAR between 10 and 19 dB.

\begin{center}
\begin{figure}
\centering
\includegraphics[width = \textwidth]{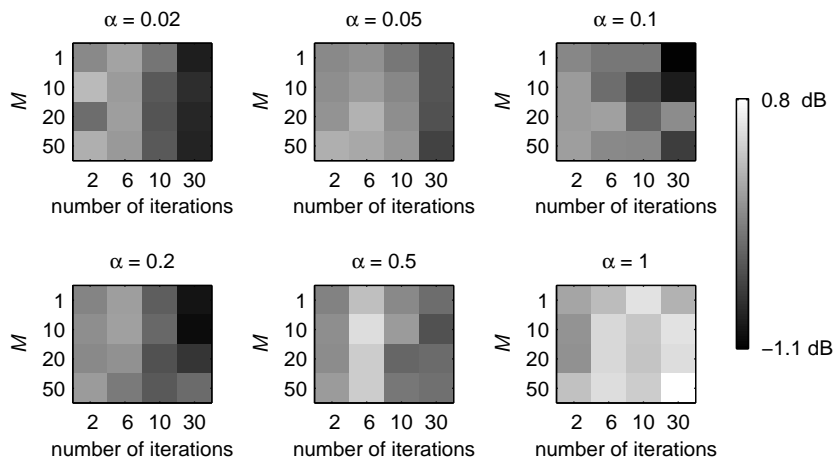}
\caption{Mean SDR for all sources and all mixtures, as a function of $\alpha$, $M$ and step size.}\label{fig:SDR}
\end{figure}
\end{center}

\begin{table}
\centering
\caption{Separation performance (dB) of the offline and best online algorithms.}\label{table:results}
\begin{tabular}{|c|c|c|c|c|c|c|c|}
\hline
Algorithm & $\alpha$ & $M$ & number of iterations & SDR & SIR & ISR & SAR \\
\hline
offline & N/A & N/A & 100 & 0.8586 & 1.2837 & 3.7989 & 13.3872\\
\hline
online & 1 & 50 & 30 & \textbf{0.8671} & 1.0675 & \textbf{3.9690} & 12.3278\\
\hline
\end{tabular}
\end{table}
As shown in table \ref{table:results}, when $\alpha = 1$, $M = 50$ and 30 GEM iterations are performed, the separation performance of the online algorithm is close to that of the offline algorithm.  For smaller block size and smaller number of iterations, the performance decreases. For example, for $M = 10$ and 6 GEM iteration, the SDR is 0.53 dB and the SIR is 3.53 dB. More generally, fig. \ref{fig:SDR} shows that for $\alpha = 1$, increasing either the block size or the number of iterations increases the SDR, though the block size has less effect on the SDR than the number of iterations. The results also show that increasing the number of iterations from 10 to 30 increases the SDR by 0.2 dB, which can be considered as a significant improvement.

When $\alpha < 1$, the SDR decreases significantly as can be seen in fig. \ref{fig:SDR}. It can also be seen that increasing the number of iterations decreases the SDR and changes of block size have little to no effect on the SDR. This can be explained by an inaccurate estimation of the model parameters of certain sources in the time intervals when these sources are inactive. These inaccurate parameters are then carried over subsequent time frames and may not converge back to accurate values. This undesirable effect is particularly salient for those parameters that are less constrained. For instance, with the considered model, the spatial covariance matrices of all sources gradually diverge towards a diffuse spatial covariance spanning all directions in the mixture, while the effect is more limited for spectral parameters which are fixed or heavily constrained.  Potential solutions to this problem are presented in the conclusion.

\section{Conclusion}

In this paper, a new framework for online audio source separation was presented. This algorithm offers an increased flexibility both in terms of the range of constraints that can be specified for each source and of the choice of a trade-off between separation accuracy and computational cost. It was shown that the separation accuracy is higher when the block size is large, but that small block sizes nevertheless offer an acceptable separation. However, small step sizes cause the spatial covariance matrices to diverge due to the presence of silence intervals in the sources.

This issue is well-known in the beamforming literature where a voice activity detector is used to restrict the time frames in which the model parameters are updated \cite{brandstein_microphone_2001}. While this solution does not readily extend to source separation, we believe that there exist a number of alternative promising solutions, e.g. adding soft constraints over the least constrained parameters by means of probabilistic priors, using different step sizes for the most constrained and the least constrained parameters, and using signal-dependent step sizes related to the power of ${\bf R}_{{\bf c}_j}(f,n)$ such that the parameters are not updated in the time intervals with low power.

Future work should also include an optimisation of the initialisation of the model parameters for each new block. After these improvements, we expect that the proposed framework will reach its full potential and provide a better trade-off between separation performance and computational cost.

\subsubsection*{Acknowledgements}
This work was supported by the EUREKA Eurostars i3DMusic project funded by Oseo.


\begin{thebibliography}{4}
\bibitem{makino_blind_2007} Makino, S., Lee, {T.-W.} and Sawada, H.: Blind Speech Separation. Springer (2007)
\bibitem{vincent_probabilistic_2010} Vincent, E., Jafari, M. G., Abdallah, S. A., Plumbley, M. D. and Davies, M. E.: Probabilistic modeling paradigms for audio source separation. In: Machine Audition: Principles, Algorithms and Systems. {IGI} Global, pp. 162--185 (2010)
\bibitem{mukai_real-time_2003} Mukai, R., Sawada, H., Araki, S. and Makino, S.: Real-time Blind Source Separation For Moving Speakers Using Blockwise {ICA} and Residual Crosstalk Subtraction. In: 4th Int. Symp. Independent Component Analysis and Blind Signal Separation, pp. 975--980 (2003)
\bibitem{mori_blind_2006} Mori, Y., Saruwatari, H., Takatani, T., Ukai, S., Shikano, K., Hiekata, T., Ikeda, Y., Hashimoto, H. and Morita, T.: Blind separation of acoustic signals combining {SIMO-model-based} independent component analysis and binary masking. {EURASIP} Journal on Advances in Signal Processing, vol. 2006, issue 1, pp. 1--17 (2006)
\bibitem{loesch_online_2009} Loesch, B. and Yang, B.: Online blind source separation based on time-frequency sparseness. In: Proc. 2009 {IEEE} Int. Conf. on Acoustics, Speech and Signal Processing, pp. 117--120 (2009)
\bibitem{togami_online_2011} Togami, M.: Online speech source separation based on maximum likelihood of local Gaussian modeling. In: Proc. 2011 {IEEE} Int. Conf. on Acoustics, Speech and Signal Processing, pp. 213--216 (2011)
\bibitem{ono_real-time_2008} Ono, N., Miyamoto, K. and Sagayama, S.: A real-time equalizer of harmonic and percussive components in music signals. In: Proc. 2008 Int. Conf. on Music Information Retrieval, pp. 139 -- 144 (2008)
\bibitem{wang_online_2011} Wang, D, Vipperla, R. and Evans, N.: Online pattern learning for non-negative convolutive sparse coding. In: Proc. Interspeech'11, pp. 65--68 (2011)
\bibitem{ozerov_general_????} Ozerov, A., Vincent, E. and Bimbot, F.: A general flexible framework for the handling of prior information in audio source separation. {IEEE} Transactions on Audio, Speech, and Language Processing, to appear
\bibitem{duong_under-determined_2010} Duong, {N.Q.K.}, Vincent, E. and Gribonval, R.: Under-determined reverberant audio source separation using a full-rank spatial covariance model. {IEEE} Transactions on Audio, Speech, and Language Processing, vol. 18, issue 7, pp. 1830--1840 (2010)
\bibitem{vincent_first_2007} Vincent, E., Sawada, H., Bofill, P., Makino, S. and Rosca, J. P.: First stereo audio source separation evaluation campaign: data, algorithms and results. In: Proc. 2007 Int. Conf. on Independent Component Analysis and Blind Source Separation, pp. 552--559 (2007)
\bibitem{brandstein_microphone_2001} Brandstein, M.S., Ward, D.B.: Microphone Arrays: Signal Processing Techniques and Applications. Springer (2001)
\end{thebibliography}
\end{document}